\documentclass[aps,prd,preprint, nofootinbib,superscriptaddress,groupedaddress]{revtex4-1}

\usepackage{amsmath}
\usepackage{amscd}
\usepackage{amsfonts}
\usepackage{amssymb}
\usepackage{mathrsfs}
\usepackage{fancybox} 
\usepackage{enumerate}
\usepackage{accents} 

\usepackage{bm}
\usepackage{amscd}
\usepackage{graphicx}
\usepackage[dvips]{color}
\usepackage{epsfig}

\makeatletter

\@addtoreset{equation}{section}
\makeatother

\usepackage{amsthm} 
\usepackage{ascmac}  

\def\half{{1\over2}}

\def\={\stackrel{\bullet}{=}}

\def\({\left(}
\def\){\right)}
\def\[{\left[}
\def\]{\right]}

\def\cM{{\cal M}}

\def\cO{{\cal O}}

\def\mbb {\mathbb }

\def \be {\begin{equation}}
\def \ee {\end{equation}}
\def \beqa {\begin{eqnarray}}
\def \eeqa {\end{eqnarray}}
\def \beal#1 {\begin{align}#1\end{align}}
\def \bes#1 {\begin{equation}\begin{split}#1\end{split}\end{equation}}

\def \nn {\nonumber}

\makeatletter

\@addtoreset{equation}{section}
\makeatother

\begin{document}

\preprint{YITP-17-102}

\title{
AdS geometry from CFT on a general conformally flat manifold 
\vspace{.3cm}
}
\author{
Sinya Aoki and Shuichi Yokoyama%
\vspace{.3cm}
}
\affiliation{ 
Center for Gravitational Physics,
Yukawa Institute for Theoretical Physics, Kyoto University,
Kitashirakawa-Oiwakecho, Sakyo-Ku, Kyoto 606-8502, Japan
\vspace{1cm} 
 }

\begin{abstract}
\vspace{.3cm}
We construct an anti-de-Sitter (AdS) geometry from a conformal field theory (CFT) defined on a general conformally flat manifold via a flow equation associated with the curved manifold, which we refer to as the primary flow equation. 
We explicitly show that the induced metric associated with the primary flow equation becomes AdS whose boundary is the curved manifold. Interestingly, it turns out that such an AdS metric with conformally flat boundary is obtained from the usual Poincare AdS by a simple bulk diffeomorphism transformation. 
We also demonstrate that the emergence of such an AdS space is guaranteed only by the conformal symmetry at boundary, which converts to the AdS isometry after quantum averaging, as in the case of the flat boundary.  

\end{abstract}

\maketitle


\section{Introduction}
\label{sec:Intro} 

The holographic principle \cite{tHooft:1993dmi,Susskind:1994vu} provides a new perspective to investigate quantum gravity on space-time with a fixed boundary. Generally in order for holography to hold in two systems, there has to exist a mechanism to invalidate extra infinite degrees of freedom which the bulk 
system usually possesses.   
In the AdS/CFT correspondence \cite{Maldacena:1997re}, which is a testable realization of holography \cite{Gubser:1998bc,Witten:1998qj}, diffeomorphism invariance clearly plays a key role to kill such extra degrees of freedom.
Thus it is important to reveal how diffeomorphism invariance is encoded in the dual field theory living on a boundary.

The AdS/CFT correspondence, due to its holographic property, can be explored by extending the dual geometry gradually from a fixed boundary and constructing the dual gravitational theory from CFT \cite{Banks:1998dd,Balasubramanian:1999ri} (see also \cite{Balasubramanian:1998sn,Balasubramanian:1998de,Bena:1999jv}). 
One natural interpretation of the emergent AdS radial direction from CFT is a conventional renormalization group (RG) scale \cite{Maldacena:1997re}.   
This interpretation was realized for a relevant RG flow from UV CFT to IR one by constructing the corresponding dual gravity solution \cite{Girardello:1998pd,Distler:1998gb,deBoer:1999tgo} (see also \cite{Skenderis:2002wp} and references therein), though a direct analysis by finding out the cut-off of Wilsonian renormalization corresponding to the sharp cut-off in the AdS radial direction is difficult to achieve due to the appearance of non-locality in the bulk \cite{Heemskerk:2010hk}.   
There were also different approaches to see a correspondence between a certain renormalization scale and the emergent AdS radial direction by using the entanglement entropy \cite{Swingle:2009bg,VanRaamsdonk:2009ar,Nozaki:2012zj}, the stochastic quantization \cite{Lifschytz:2000bj}, the bilocal field in vector models \cite{Das:2003vw}, and the flow equation 
\cite{Aoki:2015dla,Aoki:2016ohw,Aoki:2016env}. 
Recently how the Einstein equation is encoded in the boundary side was investigated \cite{Nozaki:2013vta,Bhattacharya:2013bna,Faulkner:2013ica}. 
See also \cite{Hamilton:2006az,Heemskerk:2009pn}. 

So far these analyses were almost all restricted to the asymptotic Poincare or global AdS space and payed attention to only conformal structure on the boundary \cite{FeffermanGraham,Graham:1999jg}. 
However, since the bulk theory enjoys diffeomorphism invariance, there is no reason for these analyses to be restricted on a particular AdS background. 
It should be possible for these analyses to be generalized to those on a different AdS with a more general curved boundary which admits CFT to live.  

This paper aims at making progress in this direction by using the flow equation approach \cite{Aoki:2015dla,Aoki:2016ohw,Aoki:2016env}.
A flow equation was introduced to specify how to smear operators so as to resolve a UV singularity in the coincidence limit \cite{Albanese:1987ds}, which turned out to help numerical simulation in  lattice QCD \cite{Narayanan:2006rf,Luscher:2010iy,Luscher:2009eq,Luscher:2013cpa}. 
Recently it was proposed that a one higher dimensional geometry is emergent associated with a flow equation \cite{Aoki:2015dla,Aoki:2016ohw,Aoki:2016env}
and it turned out that the direction of the free flow time precisely matches the AdS radial direction for a generic conformal field theory on the flat background \cite{Aoki:2017bru}. 
The goal of this paper is to generalize this result to an arbitrary conformally flat manifold in accord with the AdS/CFT correspondence.  
For this end we construct a flow equation for a scalar primary operator on the conformally flat background preserving the conformally symmetric structure,  
which we refer to as the {\it primary flow equation}.  
We find that the induced metric associated with the primary flow equation for a generic CFT describes an AdS space whose boundary is the  conformally flat manifold. We also show that such an AdS metric connects to the usual Poincare AdS metric by a bulk diffeomorphism transformation. 
This new result may be regarded as a consequence of the fact that the bulk theory has diffeomorphism invariance. 

The rest of this paper is organized as follows. 
In Sec.~\ref{app:CFT} we fix the setup and collect the standard technique of a conformal map to construct a CFT on a conformally flat manifold. 
In Sec.~\ref{sec:Floweq} we determine the primary flow equation.   
In Sec.~\ref{sec:InducedMetric} we compute the induced metric for this flow in a CFT on the conformally flat background, which turns out to describe the AdS space whose boundary is the conformally flat manifold. 
In Sec.~\ref{sec:Isometry} we demonstrate that the emergence of the AdS space is assured by the conformal symmetry as in the case of the flat boundary shown in \cite{Aoki:2017bru}. 
Sec.~\ref{sec:Discussion} is devoted to summary and discussion.  
In App.~\ref{app:Gmunu} we prove that the induced metric obtained in Sec.~\ref{sec:InducedMetric} describes an AdS space by explicit computation.  
In App.~\ref{app:note} we present a generic warped AdS geometry, as an example of AdS metrics with a more general curved boundary.   

\section{Conformal map of conformal field theory}
\label{app:CFT}
In this section we fix our setup in this paper and collect the standard technique to study a CFT on a conformally flat manifold by using a conformal map. 

Let us consider a real $d$-dimensional conformally flat manifold $M_d$. 
From the definition,
there exist a conformal map from a local patch $\mathbb{R}^d$ 
to a neighborhood around each point in $M_d$ such that%
\footnote{ 
For simplicity,
we consider a Riemannian manifold with the Euclidean signature in this paper.
A generalization to a different signature is straightforward by changing signs suitably.
Thus, strictly speaking, the  word ``AdS'' in the main text should read
``Euclidean AdS''.
} 
\beqa
\Omega_x = \Omega(x), \quad \Omega_x \in M_d, \quad x\in \mathbb{R}^d. 
\eeqa  
The distance in a local patch is measured by
$
ds^2 = \delta_{\mu\nu} dx^\mu dx^\nu
$ with $\mu,\nu=1,\cdots, d$, while that in the space $M_d$ is 
\beqa
(ds^2)_{M_d} &=& g_{\mu\nu}(\Omega_x) d\Omega_x^\mu d\Omega_x^\nu = 
g^{1\over d}(x) \delta_{\mu\nu} dx^\mu dx^\nu, 
\eeqa
where $g^{1\over d}(x)$ is a conformal factor associated with the curved manifold. 

Take the $d$-dimensional sphere $M_d= \mbb S^d$ with the radius $L$ as an example. 
In this case, a conformal map which covers the neighborhood around the north pole is given by a stereographic projection from the north pole $(L,0)$ to the $d$-dimensional plane $(0,\mbb R^d)$:
\beqa
\Omega^0_x &=& \frac{x^2 - 4L^2}{x^2+4L^2}L, \
\Omega_x^\mu = \frac{4L^2}{x^2+4 L^2} x^\mu,
\eeqa
where $x^2=\delta_{\mu\nu}x^\mu x^\nu$ and the sphere is embedded into $\mbb R^{d+1}$. 
The conformal factor is computed from the embedding metric as 
\beqa
g^{1\over d}(x) = \left(\frac{4L^2}{x^2+4L^2}\right)^{2}. 
\label{ConformalFactorSphere}
\eeqa

We are interested in a CFT on $M_d$ which contains a primary operator $O(\Omega_x)$ with a general conformal dimension $\Delta$. 
To construct such a CFT on $M_d$, we prepare  a CFT on a local patch $\mathbb{R}^d$ containing a primary operator $O(x)$ with the conformal dimension $\Delta$, whose conformal transformation is given by
\bes{ 
\delta^{\rm conf}O( x)
=&  (- \delta  x^\mu \partial_{ x^\mu } -{\Delta \over d} {\partial \delta x^\mu \over \partial x^\mu} )O( x),  \\
\delta x^\mu =& a^\mu + \omega^\mu{}_\nu x^\nu + \lambda x^\mu + b^\mu x^2 -2 x^\mu (b_\nu x^\nu),  
\label{ConformalTr}
}
where $a^\mu$, $\omega^{\mu\nu}$, $\lambda$ and $b^\mu$ are infinitesimal parameters of the translation, the rotation, the dilatation and the special conformal transformation, respectively.  
The scalar primary operator inserted at $\Omega_x$ is related to the one at $x$ by the pull-back of the conformal map:
\beqa
\Omega * O (\Omega_x) &:=& U_\Omega^{-1} O(\Omega_x)  U_\Omega  =
g^{-{\Delta\over 2d}}(x) O(x), 
\label{MapPrimaryOperator}
\eeqa
where $U_\Omega$ is a unitary transformation which maps a state on  $\mathbb{R}^d$ to the corresponding one on $M_d$.

Since the vacuum states on two spaces are related as
$
\vert 0 \rangle_{M_d} = U_\Omega \vert 0\rangle_{\mathbb{R}^d}
$,
correlation functions on $M_d$ are related to those on $\mathbb{R}^d$ as
\beqa  
\langle O(\Omega_{x_1}) O(\Omega_{x_2}) \cdots O(\Omega_{x_n}) \rangle_{M_d}  &=& g^{-{\Delta\over 2d}}(x_1) g^{-{\Delta\over 2d}}(x_2)\cdots g^{-{\Delta\over 2d}}(x_n) \langle O(x_1)  O(x_2) \cdots O(x_n) \rangle_{\mathbb{R}^d}  \notag
\eeqa
where $\langle \cO \rangle_{X_d} = {}_{X_d}\langle 0\vert \cO \vert 0 \rangle_{X_d}$ with $X_d=M_d,\; \mbb R^d$. 
 For example, the two point function of the scalar primary operator on $M_d$ is evaluated as
 \beqa
 \langle O(\Omega_x) O(\Omega_y) \rangle_{M_d} &=& 
g^{-{\Delta\over 2d}}(x) g^{-{\Delta\over 2d}}(y) \frac{ C }{(x-y)^{2\Delta}} =\frac{C}{\vert \Omega_x-\Omega_y\vert^{2\Delta}}, 
  \eeqa
where
$
 \vert \Omega_x-\Omega_y\vert^2 := g^{1\over 2d}(x) g^{1\over 2d}(y)(x-y)^2 
$ 
and the two point function on $\mathbb{R}^d$ is normalized as
\beqa
\langle O(x) O(y) \rangle_{\mathbb{R}^d}  =\frac{C}{(x - y)^{2\Delta}}, \quad C = { \Gamma(\Delta) \over 4^{d/2-\Delta} \pi^{{d/2}} \Gamma(d/2-\Delta) }.
\eeqa

A conformal transformation of $O(\Omega_x)$ is computed as 
\be 
\delta^{\rm conf}O(\Omega_x)
=  (- \delta \Omega^\mu_x \partial_{\Omega^\mu_x} -{\Delta \over d} {\partial \delta x^\mu \over \partial x^\mu} - \delta x^\mu \partial_\mu \log g^{\Delta \over 2d} )O(\Omega_x) 
\label{ConformalTrCurved} 
\ee
where $\delta \Omega_x^\mu = {\partial \Omega^\mu_x \over \partial x^\nu} \delta x^\nu$. 
Since correlation functions on $\mbb R^d$ are invariant under any conformal transformation, so are those on $M_d$:
\be 
\langle \delta^{\rm conf}\{ O(\Omega_{x_1}) O(\Omega_{x_2}) \cdots O(\Omega_{x_n})\} \rangle_{M_d} = 0 .
\label{ConfSymCorrelator}
\ee

\section{Primary flow equation}
\label{sec:Floweq} 

In this section we construct a certain free flow equation for a primary scalar operator on a conformally flat background.
For this purpose we begin with the one on a local patch $\mbb R^d$ \cite{Aoki:2017bru}. 
\beqa
\frac{\partial  O(x;t)}{\partial {t}} = \partial^2  O(x;t), \qquad  O(x;0) = O(x),
\label{FreeFlowFlat}
\eeqa
where $\partial^2=\delta^{\mu\nu}\partial_\mu\partial_\nu$. 
A question is how we extend this equation to the one on a conformally flat manifold $M_d$ respecting the structure of conformal property in the previous section.  

To answer this, we request a flow equation of a scalar primary operator on $M_d$ to satisfy the following properties. 
\begin{enumerate}[(i)]
 \item\label{condition1} There exists a flow time $\tilde t$ associated with $M_d$ corresponding to the flow time $t$ on $\mbb R^d$ such that the flowed operator inserted at $\Omega_x$ is related to the flowed one at $x$ by the pullback as \eqref{MapPrimaryOperator}: 
\beal{ 
\Omega *  O(\Omega_x;\tilde t) =& U_\Omega^{-1}  O(\Omega_x;\tilde t)  U_\Omega  =
g^{-{\Delta\over 2d}}(x)  O(x;t).
\label{PrimaryFlowField}
}
\item\label{condition2} The flow equation is invariant under the scale transformation.  
\end{enumerate}
We refer to a flow equation satisfying these conditions as the primary flow equation, which is determined as follows. 
The condition \eqref{condition1} fixes a differential equation consistent with \eqref{FreeFlowFlat} as
\beqa
\frac{\partial }{ \partial t}   O(\Omega_{x}; \tilde t) &=&   g^{-{\Delta\over 2d}}(x) \partial^2 g^{\Delta\over 2d}(x)  O(\Omega_{x};\tilde t), \qquad  O(\Omega_{x};0) = O(\Omega_{x}). 
\eeqa
Then $\tilde t$ is determined by the condition \eqref{condition2}, which is met if the scaling dimension of $\tilde t$ becomes two. 
Since $\tilde t$ is associated with the manifold $M_d$, namely $\tilde t$ is related to $t$ through the conformal factor, $\tilde t$ is fixed as 
\be 
\tilde t = g^{1\over d}(x) t
\label{tildet}
\ee
up to an overall constant. 
Introducing a copy of the local patch with coordinates $\tilde x^\mu$ which is independent of $\tilde t$,  
we obtain the primary flow equation as  
\beqa
\frac{\partial }{ \partial \tilde t}   O(\Omega_{\tilde x}; \tilde t) &=& D \cdot  O(\Omega_{\tilde x};\tilde t), \qquad  O(\Omega_{\tilde x};0) = O(\Omega_{\tilde x}),  
\label{FlowEquationMd}
\eeqa
with $D = g^{-{1\over d}}({\tilde x}) g^{-{\Delta\over 2d}}({\tilde x}) \tilde\partial^2 \cdot  g^{\Delta\over 2d}({\tilde x})$. 
Remark that $(x^\mu,t)$ and $(\tilde x^\mu,\tilde t)$ are two sets of independent variables, which are related by $x^\mu=\tilde x^\mu$ and \eqref{tildet}.
In what follows, however, we often abuse $\tilde x^\mu$ and $x^\mu$ just to avoid notational clutter. 

It may be instructive to mention that, in the case that $O(x)$ has the canonical dimension as $\Delta =(d-2)/2$, the operator $D$ has a conformally covariant expression such that 
\beqa
D &=& g^{\mu\nu} \nabla_\mu \nabla_\nu - \frac{d-2}{4(d-1)} R^{M_d},
\eeqa
where $g^{\mu\nu}$, $\nabla_\mu$ and $R^{M_d}$ are the metric, the covariant derivative and the scalar curvature on $M_d$, respectively. 
 
Since the two point function of the flowed primary field on a local chart is known as \cite{Aoki:2017bru}
\be 
\langle  O(x;t)  O(y;s) \rangle_{\mbb R^d} = \frac{1}{(t+s)^\Delta}F\left(\frac{(x-y)^2}{t+s}\right),
\ee
where $F$ is a smooth function depending on each CFT, that on $M_d$ is determined by using eq.~(\ref{PrimaryFlowField}) as 
\beqa
&&\langle  O(\Omega_x;\tilde t)  O(\Omega_y;\tilde s) \rangle_{M_d} =  
\frac{1}{\tilde\eta_+^\Delta} F\left(\frac{|\Omega_x-\Omega_y|^2}{\tilde \eta_+ }\right)
\eeqa
where 
\beqa
\tilde \eta_+ &:=& g^{1\over 2d}(x)g^{1\over 2d}(y)( t +  s) =
\frac{g^{1\over 2d}(y)}{g^{1\over 2d}(x)}\tilde t + \frac{g^{1\over 2d}(x)}{g^{1\over 2d}(y)}\tilde s .
\eeqa
Then the two point function of the normalized flow field 
\be 
\sigma(\Omega_x; \tilde t):= {1\over \sqrt{\langle  O(\Omega_x;\tilde t)  O(\Omega_x;\tilde t) \rangle_{M_d}}}  O(\Omega_x; \tilde t)
= \sqrt{(2\tilde t)^{\Delta} \over F(0)}  O(\Omega_x; \tilde t) 
\ee
is given by 
\beqa
\langle \sigma(\Omega_x;\tilde t) \sigma(\Omega_y;\tilde s)\rangle_{M_d} &=& 
\left(\frac{2\sqrt{\tilde t \tilde s}}{\tilde\eta_+}\right)^\Delta \bar F\left(\frac{|\Omega_x-\Omega_y|^2}{\tilde \eta_+ }\right),
\eeqa
where $\bar F(x) = F(x)/F(0)$. 

Note that the normalized flow field on $M_d$ satisfies eq.~(\ref{PrimaryFlowField}) with the vanishing conformal dimension:
\beqa
U^{-1}_\Omega\, \sigma(\Omega_x;\tilde t)\, U_\Omega &=&\Omega* \sigma(\Omega_x;\tilde t) = \sigma(x;t). 
\label{eq:sigma}
\eeqa

\section{Induced metric of CFT on a conformally flat manifold}
\label{sec:InducedMetric} 

\subsection{Induced metric}
The induced metric is defined by 
\beqa
\tilde g_{MN}(z) := R^2   \left\langle \frac{\partial \sigma(\Omega_x;\tilde t)}{\partial z^M}  \frac{\partial \sigma(\Omega_x;\tilde t)}{\partial z^N} \right\rangle_{M_d}, 
\label{DefInducedMetric}
\eeqa
where $z^M = (x^\mu, \sqrt{2d\tilde t})$, $w^M=(y^\mu,\sqrt{2d\tilde s})$ and $R$ is an arbitrary length parameter. 
Then the induced line element is given by 
\beqa
ds^2 &=& \tilde G_{M N} (Z) dZ^M dZ^N = \tilde g_{MN}(z) dz^M dz^N, 
\eeqa
where $Z^M = (\Omega^\mu_{x}, \tilde \tau)$ with $\tilde \tau = \sqrt{2 d \tilde t}$.
Explicitly the induced metric is computed as\footnote{ 
In the previous example of the $d$ dimensional sphere, $M_d =\mbb S^d$, these are computed by using \eqref{ConformalFactorSphere} as 
\bes{ 
\tilde g_{\tilde\tau\mu}(z) =& g_{\mu\tilde\tau}(z) = R^2\frac{\Delta}{\tilde\tau} \frac{2 x_\mu}{x^2+4L^2},\\
\tilde g_{\mu\nu}(z) 
=& R^2 \Delta \left(\frac{4L^2}{r^2+4L^2}\right)^2\left[ \frac{4 x_\mu x_\nu}{(4L^2)^2} + \frac{\delta_{\mu\nu}}{\tilde\tau^2}\right]. 
}
}
\bes{
\tilde g_{\tilde\tau\tilde\tau}(z) =&R^2\frac{\Delta}{\tilde\tau^2}, \\
\tilde g_{\tilde\tau\mu}(z) =& g_{\mu\tilde\tau}(z) = -R^2\frac{\Delta}{\tilde\tau} \frac{\partial}{\partial x^\mu} \log \{g^{1\over 2d}(x)\}, \label{eq:tildeg} \\
\tilde g_{\mu\nu}(z) =& R^2\Delta\left[  \frac{\partial}{\partial x^\mu}  \log \{g^{1\over 2d}(x)\}  \frac{\partial}{\partial x^\nu} \log \{g^{1\over 2d}(x)\} + \frac{\delta_{\mu\nu} g^{1\over d}(x)}{\tilde\tau^2}\right],  
}
where we use $2d\bar F^\prime(0) = - \Delta$. 
Remark that there appear nontrivial off-diagonal elements. 

An explicit calculation in App.~\ref{app:Gmunu} leads to
\beqa
G_{MN} (z) &=& - \Lambda \tilde g_{MN} (z), \qquad
\Lambda= -\frac{d(d-1)}{2R^2\Delta},
\eeqa
where $G_{MN}(z)$ is the Einstein tensor. 
As a result the induced metric turns out to be the $d+1$ dimensional (Euclidean) AdS space ($\Lambda < 0$) at $d>1$, whose radius 
is given by $R_{\rm AdS} = R \sqrt{\Delta}$.
In addition, since
\beqa
\lim_{\tilde\tau\rightarrow 0} ds^2 & =  & \frac{R^2\Delta}{\tilde\tau^2} \left( d\tilde\tau^2 + g^{1\over d}(x) \delta_{\mu\nu} dx^\mu dx^\nu \right) + O(\tilde\tau^{-1} ),
\eeqa 
the metric $\tilde g_{MN}(z)$ indeed describes the (local) AdS space in $d+1$ dimensions with the $d$ dimensional curved space $M_d$ as its boundary.\footnote{ 
The total AdS space whose boundary is $M_d$ is obtained by gluing a set of local AdS spaces whose boundaries are open covering of $M_d$, as is usual with the standard construction of a manifold. 
}

\subsection{Diffeomorphism and AdS metrics } 
In this subsection, we show that the metric $\tilde g_{MN}(z)$ is obtained from the usual Poincare AdS metric by a diffeomorphism transformation.
The AdS metric in the Poincare patch is given by
\beqa
ds^2_{\rm PAdS} &=& \frac{R^2\Delta}{\tau^2} \left(d\tau^2 +\delta_{\mu\nu} dx^\mu dx^\nu\right),
\label{PAdS}
\eeqa 
which was obtained as the result of the induced metric of a CFT on the flat space $\mbb R^d$ \cite{Aoki:2017bru}.
Under the $d+1$ dimensional diffeomorphism transformation that $(x^\mu,\tau) \to (\tilde x^\mu,\tilde\tau)$ with $\tilde x^\mu=x^\mu$ and $\tilde \tau = \tau g^{1\over 2d}(x)$, we have
\beqa
d\tau =g^{-1\over 2d}(x) \left[d\tilde\tau - \tilde\tau  \frac{\partial}{\partial x^\mu} \log \{g^{1\over 2d}(x)\} dx^\mu\right],
\eeqa 
which leads to
\beqa
ds^2_{\rm PAdS} &\to& \frac{R^2\Delta}{\tilde\tau^2} \left[ d\tilde\tau^2 - 2\tilde\tau \frac{\partial}{\partial x^\mu} \log \{g^{1\over 2d}(x)\}dx^\mu d\tilde\tau \right. \nn \\
&+&\left.
\left(\tilde\tau^2 \frac{\partial}{\partial x^\mu} \log \{g^{1\over 2d}(x)\} \frac{\partial}{\partial x^\nu} \log \{g^{1\over 2d}(x)\} + g^{1\over d}(x) \delta_{\mu\nu}\right)dx^\mu dx^\nu
\right].
\label{PAdSCFB}
\eeqa
This gives the metric identical to the one in eqs.~(\ref{eq:tildeg}).
Therefore the induced metric in the previous subsection must describe the AdS space, since it connects to the Poincare AdS by a diffeomorphism transformation, although the resulting metric has a different boundary, $M_d$.
This is a consequence of the fact that the choice of the AdS solution in the Poincare patch breaks the diffeomorphism invariance at boundary. 
  
\section{Bulk symmetry from boundary symmetry}
\label{sec:Isometry}

In this section we prove that the emergence of the AdS space is assured only by the conformal symmetry at boundary without explicit calculation. 
This was shown in the case of the flat boundary in \cite{Aoki:2017bru}. Here we argue that this is the case also for a conformally flat boundary. 

Following \cite{Aoki:2017bru} we decompose the infinitesimal conformal transformation for the normalized field $\sigma$ on a local patch derived from \eqref{ConformalTr} as  
\beqa
\delta^{\rm conf}\sigma(x;t) &=& \delta^{\rm diff}\sigma(x;t) + \delta^{\rm extra}\sigma(x;t), 
\eeqa
where 
\beqa
\delta^{{\rm diff}} \sigma(x;t)= -(\bar\delta t \partial_t + \bar\delta x^\mu \partial_\mu)\sigma (x;t), \quad 
\delta^{{\rm extra}}\sigma(x;t) 
= 4R^2 t^2 b^\mu \partial_\mu (\partial_t + {\Delta +2 \over 2t })  \sigma (x;t)~~~~~~
\label{extraterm}
\eeqa
with $ \bar\delta x^\mu =\delta x^\mu + 2d R^2 t  b^\mu,   
\bar\delta t =(2\lambda - 4 (b_\mu x^\mu) )t$. 
Then, using eq.~(\ref{eq:sigma}), we derive the conformal transformation to the normalized flow field on $M_d$ as
\beqa
\delta^{\rm conf} \sigma (\Omega_x;\tilde t)
&=& \delta^{\rm diff} \sigma(\Omega_x;\tilde t)
+ \delta^{\rm extra} \sigma(\Omega_x;\tilde t),
\eeqa 
where
\beqa
 \delta^{\rm diff} \sigma(\Omega_x;\tilde t) 
&=& (-\bar\delta \tilde t \partial_{\tilde t} -\bar\delta \Omega_x^\mu \partial_\mu) \sigma(\Omega_x;\tilde t), \\
\delta^{\rm extra} \sigma(\Omega_x;\tilde t) &=& 
{4g^{-{1\over d}}(x)\tilde t^2 b^\mu(\partial_\mu\log\{ g^{1\over d}(x)\}+\partial_\mu) }\left(\partial_{\tilde t} +\frac{\Delta+2}{2\tilde t}\right) \sigma(\Omega_x;\tilde t),
\eeqa
with $\bar\delta \tilde t = \bar\delta t\, g^{ 1\over d}(x) +\bar\delta x^\nu\, \tilde t\,  \partial_\nu \log\{ g^{1\over d}(x)\}, \; \bar\delta \Omega_x^\mu = \bar\delta x^\nu \frac{\partial \Omega_x^\mu}{\partial x^\nu}.$

Let us show that the induced metric \eqref{DefInducedMetric} is invariant under the transformation $\delta^{\rm diff}$:
\beqa
\delta^{\rm diff} \tilde g_{MN}(z)
&=& \lim_{w\to z}\frac{\partial}{\partial z^M}\frac{\partial}{\partial w^N}
(\delta^{\rm conf} -\delta^{\rm extra} ) 
\langle \sigma(\Omega_x;\tilde t) \sigma(\Omega_y;\tilde s)
\rangle_{M_d}  \nn \\
&=& - \lim_{w\to z}\frac{\partial}{\partial z^M}\frac{\partial}{\partial w^N}\delta^{\rm extra}
\langle \sigma(\Omega_x;\tilde t) \sigma(\Omega_y;\tilde s)
\rangle_{M_d},
\label{eq:diff}
\eeqa
where we used the conformal symmetry of correlation functions \eqref{ConfSymCorrelator}:
\beqa
\delta^{\rm conf} \langle \sigma(\Omega_x;\tilde t) \sigma(\Omega_y;\tilde s)
\rangle_{M_d} 
= 0. \nn 
\eeqa
Since 
\beqa
\delta^{\rm extra}
\langle \sigma(\Omega_x;\tilde t) \sigma(\Omega_y;\tilde s)
\rangle_{M_d} 
&=&- 8\frac{\sqrt{4\tilde t\tilde s}^\Delta}{(\tilde \eta_+)^\Delta}\left(\frac{\tilde t}{g^{1\over d}(x)}-\frac{\tilde s}{g^{1\over d}(y)}\right)b_\mu (x-y)^\mu (x-y)^2 F^{\prime\prime}\left(\frac{|\Omega_x-\Omega_y|^2}{\tilde\eta_+}\right), \nn
\eeqa
which vanishes in the $w\to z$ limit, so does eq.~(\ref{eq:diff}).
Explicitly
\beqa
\delta^{\rm diff}  \tilde g_{MN}(z) &=& -\bar\delta  z^K\partial_K \tilde g_{MN}(z) -
\partial_M\bar\delta  z^K \tilde g_{KN}(z) -  \partial_N\bar\delta  z^K \tilde g_{MK}(z) = 0,
\eeqa
with $\bar\delta x^\mu =\delta x^\mu + g^{-{1\over d}}(x) \tilde\tau^2 b^\mu, \,\bar\delta \tilde \tau = \tilde\tau\left[\lambda - 2\, (b\cdot x)  + \bar\delta x^\mu\, \partial_\mu \log \{g^{1\over 2d}(x)\} \right]$. 
This means that the induced metric is invariant under the infinitesimal AdS isometry transformations expressed in the coordinates $(x^\mu, \tilde\tau)$. 
As a result, the induced metric has to be the AdS one since it is a maximally symmetric space. This completes the proof of our claim. 

\section{Discussion} 
\label{sec:Discussion}

In this paper, we have extended our previous investigation \cite{Aoki:2017bru} on the proposal \cite{Aoki:2015dla,Aoki:2016ohw, Aoki:2016env} that the bulk geometry is constructed from a boundary CFT on the flat background to the case with an arbitrary boundary CFT on curved spaces within a conformally flat class by using a canonical flow equation called the primary one.
We have shown that the resulting induced metric becomes AdS whose boundary is the curved manifold. We have found that such AdS metrics can be constructed from the usual Poincare AdS metric by a simple diffeomorphism transformation to the AdS radial direction with the other directions fixed. 
We have also shown that the conformal symmetry at the boundary generates the AdS isometry for the vacuum expectation value of the metric operator, so that the bulk geometry must be AdS with a given boundary, as in the case of the flat boundary.
In App.~\ref{app:note}, a generic warped AdS geometry with a curved boundary was presented as an example of a different class of AdS geometries.

In the previous paper \cite{Aoki:2017bru}, we had shown that the induced metric associated with the free flow equation from the CFT on the flat boundary describes the usual Poincare AdS space as \eqref{PAdS}.
Let us recast this in the following way.
Since the flowed field $\sigma$ is dimensionless thank to the (NLSM) normalization and free from the UV divergence at non-zero flow time, 
the Poincare symmetry as well as an absence of dimensionful constants in a CFT on the flat boundary demand the metric to be of a form such that $ds_{d+1}^2 = R^2 \{ A({x^2 / \tau^2}) d\tau^2 + B({x^2 / \tau^2})  \delta_{\mu\nu} dx^\mu dx^\nu \}/\tau^2$.
We had shown that the symmetry argument constrains the functions $A$ and $B$ to be an equal constant while an explicit calculation determines its value as $\Delta$, so that the induced metric results in the Poincare AdS space.
This kind of analysis, however, may not be available if the boundary space is curved.
Fist of all, it is not guaranteed that there exists the AdS metric with a curved boundary in the Poincare patch.
Secondly, extra dimensionful parameters such as the radius of the sphere existing in a curved space make the ansatz of the metric much more complicated.
Regardless of these difficulties,  
the method proposed in Refs.~\cite{Aoki:2015dla,Aoki:2016ohw, Aoki:2016env} still works well for a CFT on a conformally flat boundary such as a $d$ dimensional sphere. The induced metric for such a CFT was explicitly given by \eqref{PAdSCFB}, 
which was shown to describe the AdS space with the boundary metric $g^{1\over d}(x) \delta_{\mu\nu}$. To the best of our knowledge, construction of an AdS metric with a general conformally flat boundary was not known, or, at least our result \eqref{PAdSCFB} gives a new expression for such a non-trivial AdS space with the manifest Poincare symmetry. 

Although it is interesting to generalize our analysis to the case with a more general curved boundary including de-Sitter space or a gravity solution corresponding to a RG flow on a curved manifold, 
one of the most important and urgent issues in the proposal \cite{Aoki:2015dla,Aoki:2016ohw, Aoki:2016env} is to clarify how this formalism encodes the bulk dynamics beyond the geometry.
For this purpose, the equation of motion for (the fluctuation of) the metric must be determined by calculating, for instance, the 2-point correlation function of the metric operator $g_{MN}(z)$.
Works along this direction are ongoing.

\section*{Acknowledgement}
S. A. is supported in part by the Grant-in-Aid of the Japanese Ministry of Education, Sciences and Technology, Sports and Culture (MEXT) for Scientific Research (No. JP16H03978),  
by a priority issue (Elucidation of the fundamental laws and evolution of the universe) to be tackled by using Post ``K" Computer, 
and by Joint Institute for Computational Fundamental Science (JICFuS).

\appendix 
\section{Explicit calculation of the Einstein tensor}
\label{app:Gmunu}
In this appendix we explicit compute the Einstein tensor for the induced metric \eqref{eq:tildeg} to result in the (Euclidean) AdS space. 
The metric and its inverse are given by
\beqa
\tilde g_{\tilde\tau\tilde\tau}(z) &=& \frac{R_{\rm AdS}^2}{\tilde\tau^2}, \quad \tilde g_{\tilde\tau\mu}(z) = \tilde g_{\mu\tilde\tau}(z) = -\frac{R_{\rm AdS}^2}{\tilde\tau} F_\mu(x), \quad
\tilde g_{\mu\nu}(z) = R_{\rm AdS}^2\left[ F_\mu(x) F_\nu(x) + \frac{\delta_{\mu\nu} \Omega^2(x)}{\tilde\tau^2}\right],\nn \\ 
\tilde g^{\tilde\tau\tilde\tau}(z) &=& \frac{\tilde\tau^2}{R_{\rm AdS}^2}\left[1+\tau^2 F^2(x)\right], \quad
\tilde g^{\tilde\tau\mu}(z) = \tilde g^{\mu\tilde\tau}(z) = \frac{\tilde \tau \tau^2}{R_{\rm AdS}^2} F^\mu, \quad
\tilde g^{\mu\nu}(z) = \frac{\tau^2}{R_{\rm AdS}^2}\delta^{\mu\nu}, 
\eeqa
where $R_{\rm AdS}^2= R^2\Delta$, $\Omega(x) = g^{1\over 2d}(x)$, 
\beqa
F_{\mu}(x) &=& \frac{\partial}{\partial x^\mu} \log \Omega(x), \quad
F^\mu(x)=\delta^{\mu\nu} F_\nu(x), \quad F^2(x) = F^\mu(x) F_\mu(x) .
\eeqa
We have
\beqa
\partial_{\tilde\tau} \tilde g_{\tilde\tau\tilde\tau} &=& -2\frac{R_{\rm AdS}^2}{\tilde \tau^3}, \quad
\partial_\mu \tilde g_{\tilde\tau\tilde\tau} = 0, \quad
\partial_{\tilde\tau}\tilde g_{\tilde\tau\mu} = \frac{R_{\rm AdS}^2}{\tilde\tau^2} F_\mu, \quad
\partial_{\nu}\tilde g_{\tilde\tau\mu} = -\frac{R_{\rm AdS}^2}{\tilde\tau}\partial_\nu F_\mu, \nn \\
\partial_{\tilde\tau} \tilde g_{\mu\nu}&=& -2\frac{R_{\rm AdS}^2 \Omega^2}{\tilde\tau^3} \delta_{\mu\nu}, \quad
\partial_{\alpha} \tilde g_{\mu\nu} = R_{\rm AdS}^2\left[ \partial_\alpha ( F_\mu F_\nu) + 2 \frac{F_\alpha}{\tau^2}\delta_{\mu\nu}\right] .
\eeqa

Christoffel symbols,
\beqa
\Gamma_{LM}^K = \frac{1}{2}\tilde g^{KN}\left[ \partial_M \tilde g_{NL} +  \partial_L \tilde g_{NM} -  \partial_N \tilde g_{LM}\right],
\eeqa 
is evaluated as
\beqa
\Gamma^{\tilde\tau}_{\tilde\tau\tilde\tau} &=& -\frac{1}{\tilde\tau}, \quad
\Gamma^{\mu}_{\tilde\tau\tilde\tau} = 0, \quad
\Gamma^{\tilde\tau}_{\tilde\tau\mu}= -F_\mu, \quad
\Gamma^{\tilde\tau}_{\mu\nu}= \frac{\Omega^2}{\tilde\tau}\delta_{\mu\nu} + 3\tilde\tau F_\mu F_\nu
-\tilde\tau F_{\mu\nu}, \nn \\
\Gamma^{\mu}_{\tilde\tau\nu}&=& -\frac{1}{\tilde\tau}\delta^\mu_\nu, \quad
\Gamma^{\alpha}_{\mu\nu}= F_\nu \delta^\alpha_\mu + F_\mu\delta^\alpha_\nu ,
\eeqa
where $
F_{\mu\nu}=F_{\nu\mu} := \frac{\partial_\mu\partial_\nu \Omega}{\Omega}.
$
Then the Ricci curvature,
\beqa
R_{LM} &=& \partial_N \Gamma^N_{ML} - \partial_M \Gamma^N_{NL} + \Gamma^N_{NK}\Gamma^K_{ML} - \Gamma^N_{MK}\Gamma^K_{NL}, 
\eeqa 
is given by
\beqa
R_{\tilde\tau\tilde\tau} &=& -\frac{d}{\tilde\tau^2}, \quad
R_{\tilde\tau\nu}=R_{\nu\tilde\tau} =\frac{d}{\tilde \tau} F_\nu, \quad
R_{\mu\nu} = -d\left[ F_\mu F_\nu +\frac{\Omega^2}{\tilde\tau^2}\delta_{\mu\nu} \right],
\eeqa
which can be written as
\beqa
R_{LM} = -\frac{d}{R_{\rm AdS}^2} \tilde g_{LM}  \rightarrow R := \tilde g^{LM} R_{LM} = -\frac{d(d+1)}{R_{\rm AdS}^2}.
\eeqa
Therefore the Einstein tensor, $G_{LM}:=R_{LM} - \frac{1}{2}\tilde g_{LM} R$, satisfies 
\beqa
G_{MN}(z) = - \Lambda \, \tilde g_{MN}(z), \qquad \Lambda =-\frac{d(d-1)}{2 R^2\Delta}.
\eeqa

\section{General warped AdS geometry} 
\label{app:note} 

In this appendix we give another example of AdS metric whose boundary is not necessarily conformally flat.
For this purpose we consider a generic warped geometry with one warped dimension, whose metric is written as   
\be 
ds'^2 = f_1(\tau) d\tau^2 + f_2(\tau) ds^2_\cM, 
\label{anotherAdS}
\ee
where $\cM$ is a $d$ dimensional manifold whose metric $ds^2_\cM = g_{\mu\nu} dx^\mu dx^\nu$ independent of $\tau$. 
We claim that this warped geometry can become AdS by a specific choice of $f_{1,2}(\tau)$
if and only if the base manifold $\cM$ is an Einstein manifold, namely the (pseudo-)Riemannian manifold 
whose Ricci tensor is constantly proportional to its metric. 
 The proof of this statement is given as follows. 

The condition for the AdS$_{d+1}$ is given by
\be 
G'_{MN} + g'_{MN} \Lambda = 0,
\ee
where $G'_{MN} = R'_{MN} - \half g'_{MN}R'$, $\Lambda < 0$ and 
\be 
g'_{\tau\tau} = f_1, \quad g'_{\mu\nu} = f_2 g_{\mu\nu}, \quad g'_{\mu\tau} = 0
\ee 

We first perform the Weyl transformation so that 
\be 
ds^2 = e^{-2\phi} ds'^2 =N^2 d\tau^2 + ds^2_\cM = g_{MN} dx^M dx^N, 
\ee
where $\phi = \half \log f_2(\tau)$, $N^2 = f_1(\tau)/f_2(\tau)$. 
The Ricci tensor becomes
\bes{ 
R'_{NL} =& R_{NL} - g_{NL} \partial^2\phi + (d-1) ( -\nabla_N\partial_L\phi + \partial_N\phi\partial_L\phi-g_{NL}\partial_M\phi\partial^M\phi), \\
R'=& e^{-2\phi}( R - (2d-1) \partial^2\phi - d(d-1)\partial_M\phi\partial^M\phi).
}
Since the function $N$ can be absorbed in the $\tau$ coordinate by diffeomorphism, there is no non-trivial curvature in $\tau$ direction. 
Thus 
\bes{
R_{\mu\nu} =& R^\cM_{\mu\nu}, \quad R= R^\cM, \quad 
R_{\mu\tau} = R_{\tau\tau}= 0.  
\label{Ricci}
}
Therefore the Ricci tensor is computed as 
\bes{ 
R'_{\mu\nu} 
=& R_{\mu\nu} - g_{\mu\nu} \partial^2\phi + (d-1) (-g_{\mu\nu} g^{\tau\tau} \partial_\tau\phi\partial_\tau \phi), \\
R'_{\tau\nu} =& R_{\tau\nu},  \\
R'_{\tau\tau} 
=& R_{\tau\tau} - g_{\tau\tau} \partial^2\phi + (d-1) (-\nabla_\tau\partial_\tau\phi),\\
R'=& e^{-2\phi}( R - (2d-1) \partial^2\phi  -  d(d-1)g^{\tau\tau}\partial_\tau\phi\partial_\tau\phi). 
}
Thus the Einstein tensor is given as
\bes{ 
G'_{\mu\nu} 
=& G_{\mu\nu} +(d-1) g_{\mu\nu} \partial^2\phi + \half(d-1)(d-2) g_{\mu\nu} g^{\tau\tau} \partial_\tau\phi\partial_\tau \phi, \\
G'_{\tau\nu} =& R_{\tau\nu},  \\
G'_{\tau\tau} 
=&G_{\tau\tau} + \half d(d-1) \partial_\tau\phi\partial_\tau \phi.
}
By using \eqref{Ricci}, the AdS condition becomes 
\bes{  
& G^\cM_{\mu\nu} +(d-1) g_{\mu\nu} \partial^2\phi + \half(d-1)(d-2) g_{\mu\nu} g^{\tau\tau} \partial_\tau\phi\partial_\tau \phi  +f_2 g_{\mu\nu} \Lambda = 0,  \\
& R_{\tau\nu} = 0, \\
& -\half g_{\tau\tau} R^\cM  + \half d(d-1)  \partial_\tau\phi\partial_\tau \phi +f_1 \Lambda = 0 .
}
The 3rd equation gives  
\be 
f_1 = { -{1\over 8} d(d-1)  (f_2'/f_2)^2 \over \Lambda - R^\cM / (2f_2)}.
\label{f1}
\ee
Since $f_1$ is dependent only on $\tau$, $R^\cM$ has to be a constant. 
Plugging this into the 1st equation leads to
\be
R^\cM_{\mu\nu} -{1\over d} g_{\mu\nu}R^\cM +(d-1) g_{\mu\nu} \partial^2\phi  +{2 \over d} f_2 g_{\mu\nu} \Lambda = 0.
\label{1st}
\ee
Since \eqref{f1} implies
\be
(d-1) g_{\mu\nu} \partial^2\phi  +{2 \over d} f_2 g_{\mu\nu} \Lambda = 0, 
\label{f2}
\ee
 \eqref{1st} becomes
\be
R^\cM_{\mu\nu} -{1\over d} g_{\mu\nu}R^\cM = 0,
\label{EinsteinMfd}
\ee
which, together with the constant $R^\cM$, means that $\cM$ is an Einstein manifold.
This completes the proof of our claim. 

For example, if we take $f_2(\tau) = c_0/\tau^2$ with a constant $c_0$, we have
\be
f_1(\tau) = \frac{c_0}{\tau^2}\left[\frac{2d(d-1)}{-c_0\Lambda + \tau^2 R^\cM/2}\right].
\ee

\bibliography{CFT2AdSCurvedMfd}

\end{document}